\documentclass[prl,twocolumns,floatfix,showpacs,nofootinbib]{revtex4}
\usepackage[T2A]{fontenc}
\usepackage[cp1251]{inputenc}
\usepackage{graphicx}
\usepackage{epsf}
\usepackage{epsfig}
\usepackage{amssymb,amsthm,amsmath}
\usepackage{latexsym}

\begin{document}

\preprint{}
\title{  Spatial image of reaction area from scattering.
II
:\\
{\small On connection between the differential cross-sections in
transverse
   momentum and in nearest approach parameter.}}

\author{N.Bobrovskaya}
   \affiliation{Department of Theoretical Physics, Irkutsk State University, Irkutsk, 664003
  Russia}

\author{M.V.Polyakov}%
 \email{Maxim.Polyakov@tp2.ruhr-uni-bochum.de}
  \affiliation{Institut fur Theoretische Physik II
   Ruhr-Universitaet Bochum, NB6 D-44780 Bochum, Germany}

\author{A.N.Vall}
 \email{vall@irk.ru}
  \affiliation{Department of Theoretical Physics, Irkutsk State University, Irkutsk, 664003
  Russia}
\author{A.A.Vladimirov}
 \email{avlad@theor.jinr.ru}
  \affiliation{
  Bogoliubov Laboratory of Theoretical Physics,
  JINR, 141980, Moscow Region, Dubna, Russia}


%
%
%
%

\begin{abstract}
The connection between differential cross sections of particle $C$
creation on transverse momentum and on nearest approach parameter
$\vec{b}$ is investigated in the context of formalism $SO(2,1)$
algebra. Where parameter $\vec b$ characterizes particle creation
area. This distribution tightly concerned with spatial structure
of particles interaction and allows intuitive physic
interpretation. It is shown that area of large transverse momentum
carries in the main contribution to distribution function on $b$
in backward semisphere in interval $1/2 \leq bq \sim 1 $, where
$q$ is momentum of the particle $C$. The left border of inequation
defines by Geizenberg uncertainty relation on parameters "momentum
- radius of localization area" . The spatial structure of creation
area in transverse momentum plane is a set of discrete
axially-symmetrical (for spinless particles $C$) zones. The
received connection between cross sections on $\vec{q}_{\bot}$ and
$\vec{b}$ is exact and don't connect with any model. So ability
appears to analyze the spatial structure of target using
experimental data of particle $C$ angle distribution. It was
received an exact relation between $<b^2_\pm>$ and
$<\cos^2\theta_\pm>$ for any $A+B\rightarrow C+D$ processes in
center of particles $A$ and $B$ mass frame, where average is going
on corresponding differential cross-sections. It is shown that
quantum-mechanic constrain on $b^2$ spectrum ($b^2q^2>\hbar^2/4$)
brings to constrain $<\cos^2\theta_\pm>\geqslant1/8$. As
application it is considered the process $\gamma + p \rightarrow
\pi ^0 + p$ at photon energy $ E_{\gamma} = 5 \textrm{Gev}$.
\end{abstract}

\maketitle

\section{Differential cross-section on the transverse momentum of particle $C$}
In this paper we will reproduce in detail the derivation of
connection between the differential cross-section in particle $C$
momentum and $S$-matrix element \cite{Shirkov}. As in the
framework of the same scheme the similar calculation for the
cross-section on  nearest approach parameter $b$ \cite{part1} will
be made.

Let us examine the process  $ A+B\rightarrow C+D $\ for two
physical cases. In the first case particle $C$ is created with
fixed momentum $\vec{q}$ \ , the second, particle $C$ is created
in state $|\vec{\mu} ,q,\epsilon> $. The differential
cross-section on transverse momentum of the particle $C$ is
\begin{equation}\label{diff1}
\frac{d\sigma ^{\pm}}{\mathrm{d}\Omega_{\vec{q}}}= \frac{1}{n_1
n_2 T V |\vec{u}|}\frac{dN ^{\pm}}{\mathrm{d}\Omega_{\vec{q}}} ~.
\end{equation}
Here $\vec{u}$ is relative speed of initial particles, $n_1 ,n_2\
$ are densities of particles $A$ and $B$, $(\pm)$ means that the
particle $C$ scatters to the forward and backward half-sphere
($z$-axis is directed along the momentum of particle $A$ ), and
$N$ is the total number of particles $C$ and $D$ creation events
in all space (volume $V$) during infinite time ($T$). This is
\begin{equation}\label{diff2}
\begin{split}
 N&=\int \mathrm{d}\vec{q} \ \mathrm{d}\vec{q}_{1}\  |< \vec{q} ;\vec{q}_{1}|\hat{F} |in  > |^2 =
 \sum \limits_{\epsilon=\pm 1}N^{(\epsilon )} =\\
 &=\sum \limits_{\epsilon=\pm 1}\int \mathrm{d}\vec{q}_{1}\ q^2 dq \
   \mathrm{d}\Omega_{\vec{q}} \ |<\vec{q}_{\bot},\epsilon \sqrt{q^2 -q_{\bot}^{2}};\vec{q}_{1}|\ \hat{F}\
   |in>|^2~,
 \end{split}
\end{equation}
where
$$|in>=(2\pi)^{3}n^{1/2}_1 n^{1/2}_2a^{+}(\vec{p}_1)\ a^{+}(\vec{p}_2)\ |0> \equiv (2\pi)^{3}n^{1/2}_1 n^{1/2}_2|\vec p_1 ; \vec p_2>~. $$
The creation operators are related to particles $A$ and $B$ ,
operator $\hat{F}$ is connected with $S$-matrix by relation
$S=I+i\hat{F}$. Translational invariance allows present the matrix
element in the form:
\begin{equation}\label{diff2a}
 <f|\hat{F}|in> = \delta^{(4)}(q_{in}-q_{f})<f|\hat{A}|in>~,
\end{equation}
therefore:
\begin{equation}\label{diff3}
\begin{split}
  N&=\frac{T V}{(2\pi)^4} \sum \limits_{\epsilon=\pm 1}\int \mathrm{d}\vec{q}_{1}\ q^2 dq \
   \mathrm{d}\Omega_{\vec{q}} \ \delta^{(4)}(q+q_1 -P)\ \times \\
   &\times |<\vec{q}_{\bot},\epsilon \sqrt{q^2 -q_{\bot}^{2}};\vec{q}_{1}|\ \hat{A}\ |in>|^2 =\\
   &= \frac{T V}{(2\pi)^4} \sum \limits_{\epsilon=\pm 1}\int  q^2 dq \
   \mathrm{d}\Omega_{\vec{q}} \ \delta(E_{q}+\sqrt{(\vec{q}-\vec{P})^2+m_{D}^{2}} -P^0)  \times \\
   &\times |<\vec{q}_{\bot},\epsilon \sqrt{q^2 -q_{\bot}^{2}};\vec{q}_{1}=\vec{P}-\vec{q}\ |\ \hat{A}\
   |in>|^2~.
\end{split}
\end{equation}
Here, the four-momentum $P=P_1+P_2$. In (\ref{diff3}) the $\delta$
- function determines the energy shell for particle $C$ reaction :
\begin{equation}\label{diff4}
  \left \{ E_{q}+\sqrt{(\vec{q}-\vec{P})^2+m_{D}^{2}} -P^0)   \right
  \}_{q=\tilde{q}}=0~.
\end{equation}
Let us define:
\begin{equation}\label{diff5}
\begin{split}
E_{D}&=\sqrt{(\vec{q}-\vec{P})^2+m_{D}^{2}}\\
\lambda (\vec{q}_{\bot} )&=\left \{\frac{q E_{q} E_{D}}{q^2 P^0
-(\vec{q}\cdot \vec{P})E_{q}}\right \}_{q=\tilde{q}}~.
\end{split}
\end{equation}
Then the following relation is valid:
\begin{equation}\label{diff6}
  \delta(E_{q}+\sqrt{(\vec{q}-\vec{P})^2+m_{D}^{2}} -P^0)=\lambda (\vec{q}_{\bot} )\
  \delta(q-\tilde{q})~.
\end{equation}
Integrating (\ref{diff3}) over $q$ we get:
\begin{equation}\label{diff7}
\begin{split}
  N&= \frac{T V}{(2\pi)^4} \sum \limits_{\epsilon=\pm 1}\int  \mathrm{d}\Omega_{\vec{q}} \ q^2\ \lambda (\vec{q}_{\bot} )\times \\
  \times &\ |<\vec{q}_{\bot},\epsilon \sqrt{q^2 -q_{\bot}^{2}};\vec{q}_{1}=\vec{P}-\vec{q}\ |\ A\
  |in>|^2\ \ ,\ \ q=\tilde{q}\ \ ,\ \ q=|\vec{q}|~.
  \end{split}
\end{equation}
So, we obtain a relation for particle $C$ transverse momentum
distribution:
\begin{equation}\label{diff8}
  \frac{dN ^{\pm}}{\mathrm{d}\Omega_{\vec{q}}}=\frac{T V}{(2\pi)^4}  \ q^2 \lambda (\vec{q}_{\bot} )\ \
  |<\vec{q}_{\bot},\pm \sqrt{q^2 -q_{\bot}^{2}};\vec{q}_{1}=\vec{P}-\vec{q}\ |\ A\
  |in>|^2\\ \ ,\ \ q=\tilde{q}~.
\end{equation}
Let us define:
\begin{equation}\label{14a}
  A^{(\epsilon )}(\vec{q})=<\vec{q}_{\bot},\epsilon \sqrt{q^2 -q_{\bot}^{2}};\vec{q}_{1}=\vec{P}-\vec{q}\  |\ A\
  |\vec p_1 ; \vec p_2>\ ,\ \ q=\tilde{q}~.
\end{equation}
Then (\ref{diff8}) takes a view:
\begin{equation}\label{diff8a}
  \frac{dN ^{\pm}}{\mathrm{d}\Omega_{\vec{q}}}=(2\pi)^2T V n_1 n_2  \ q^2 \lambda (\vec{q}_{\bot} )\
  |A^{(\pm )}(\vec{q})|^2\\ \ ,\ \ q=\tilde{q}~.
\end{equation}
The differential cross-section is resulted  by substitution
(\ref{diff8a}) to relation (\ref{diff1}).

Here we mark an important notice. An angle dependents in cross
section $\frac{d \sigma ^{\pm}}{\mathrm{d}\Omega_{\vec{q}}}$
appears not only through variable $\vec{q}_{\bot}$, but in general
case also through $\tilde{q}$. In the frame of particles $A$ and
$B$ centre of mass (c.m.f.) the particle $C$ energy defined only
through square of invariant mass $s=(p_{A}+ p_{B})^2$. And it is
\begin{equation}\label{Escm}
   E_{C}^{\star}= \frac{1}{2\sqrt{s}} (s + m_{C}^2 -m_{D}^2)~.
\end{equation}
So:
 $$ \tilde{q} =p_{C}^{\star}=\sqrt{(E_{C}^{\star})^2 -m_{C}^2}~. $$
And $E_C^\star$ isn't depend for scattering angle (the star means
c.m.f.). The situation changes in the labor frame (l.f.). In this
case there are critical parameter $\kappa$ which is
\cite{kinematica}
 $$ \kappa = \frac{\sqrt{s}\cdot p_{C}^{\star}}{m_{C}\cdot p_{A}}
 $$~.
In the case $\kappa>1$ the connection between particle $C$
momentum and scattering angle has a single meaning and has a view:
\begin{equation}\label{L.S.C.}
\tilde{q}=p_{C}= \frac{\sqrt{s}E_{C}^{\star}p_{A}\cos (\theta) +
(E_{A}+m_{A})[s(p_{C}^{\star})^2 -m_{C}^2 p_{A}^2 \sin^2 (\theta )
]^{1/2}}{(E_{A}+m_{B})^2 -p_{A}^2 \cos^2 (\theta )}~,
\end{equation}
where $0<\theta<\pi $ is scattering angle in l.f. (there isn't the
maximum angle limitation). We will back to discuss this expression
below.

\section{The differential cross-section on  nearest approach parameter $\vec b$ in c.m.f.}
Now we discuss the case, when the particle $C$ creates in state
$|\vec{\mu } ,q,\epsilon > $ . Choosing for state $|out>$ the
corresponding basis, we obtain following relation for the total
number of $C$ and $D$ particles creation events \cite{part1}:
\begin{equation}\label{diff9}
  N=(2\pi)^2 \sum \limits_{\epsilon=\pm 1}\int \mathrm{d}\vec{q}_{1}\ q^2 dq \
   \mathrm{d}\Omega_{\vec{\mu}} \ |<\ \vec{\mu} ,q,\epsilon ; \vec{q}_{1}\ |\ \hat{F}\
   |in>|^2~.
\end{equation}

As distinct from the previous case, the state $<\ \vec{\mu}
,q,\epsilon ; \vec{q}_{1}\ |$ is not a state with a define
momentum. So the  particle $C$ cross-section in this state can't
be expressed through the matrix element $<\ \vec{\mu} ,q,\epsilon
; \vec{q}_{1}\ |\ \hat{F}\ |in>$. To find $\frac{d\sigma
^{\pm}}{\mathrm{d}\Omega_{\vec{\mu}}}$ we will use a state
expansion on states with the fixed particle $C$ transverse
momentum \cite{part1}:
\begin{equation}\label{diff10}
 <\ \vec{\mu } ,q,\epsilon ; \vec{q}_{1}\ | =\frac{1}{(2 \pi)^2}\int \bar{\xi }(\vec{q}_{\bot},\vec{\mu })
 <\vec{q}_{\bot},\epsilon \sqrt{q^2 -q_{\bot}^{2}};\vec{q}_{1}|\
 \mathrm{d}\Omega_{\vec{q}}~.
\end{equation}
Substituting this expansion to (\ref{diff9}) we obtain:
\begin{equation}\label{diff11}
\begin{split}
N&=\frac{1}{(2\pi)^2}\sum \limits_{\epsilon=\pm 1}\int \mathrm{d}\vec{q}_{1}\ q^2 dq \
   \mathrm{d}\Omega_{\vec{\mu}}\  \mathrm{d}\Omega_{\vec{q}} \
   \mathrm{d}\Omega_{\vec{k}}\ \bar{\xi }(\vec{q}_{\bot},\vec{\mu })\ \xi (\vec{k}_{\bot},\vec{\mu })\times \\
   \times &  F(\vec{q}_{\bot},\vec{q}_{1})\ \bar{F}(\vec{k}_{\bot},\vec{q}_{1}) \ \ ,\
   |\vec{q}|=|\vec{k}|=q~,
\end{split}
\end{equation}
where we use a notation:
\begin{equation}\label{diff11a}
  <\vec{q}_{\bot},\epsilon \sqrt{q^2 -q_{\bot}^{2}};\vec{q}_{1} |\ \hat{F}\
  |in>=F(\vec{q}_{\bot},\vec{q}_{1})~.
\end{equation}
Let us make an identical transformation in (\ref{diff11}):
$$F(\vec{q}_{\bot},\vec{q}_{1})=[F(\vec{q}_{\bot},\vec{q}_{1})-F(\vec{k}_{\bot},\vec{q}_{1})]+F(\vec{k}_{\bot},\vec{q}_{1})~,$$
$$\bar{F}(\vec{k}_{\bot},\vec{q}_{1})=[\bar{F}(\vec{k}_{\bot},\vec{q}_{1})-\bar{F}(\vec{q}_{\bot},\vec{q}_{1})]+\bar{F}(\vec{q}_{\bot},\vec{q}_{1})~.$$
Then:
\begin{equation}\label{diff11b}
\begin{split}
N&=\frac{1}{(2\pi)^2} Re
 \sum \limits_{\epsilon=\pm 1}\int \mathrm{d}\vec{q}_{1}\ q^2 dq \
   \mathrm{d}\Omega_{\vec{\mu}}\  \mathrm{d}\Omega_{\vec{q}} \
   \mathrm{d}\Omega_{\vec{k}}\ \bar{\xi }(\vec{q}_{\bot},\vec{\mu })\ \xi (\vec{k}_{\bot},\vec{\mu
   })|F(\vec{q}_{\bot},\vec{q}_{1})|^2- \\
   &-\frac{1}{(2\pi)^2}\frac{1}{2}\sum \limits_{\epsilon=\pm 1}\int \mathrm{d}\vec{q}_{1}\ q^2 dq \
   \mathrm{d}\Omega_{\vec{\mu}}\  \mathrm{d}\Omega_{\vec{q}} \
   \mathrm{d}\Omega_{\vec{k}}\ \bar{\xi }(\vec{q}_{\bot},\vec{\mu })\ \xi (\vec{k}_{\bot},\vec{\mu })\times \\
   \times &  |F(\vec{q}_{\bot},\vec{q}_{1})\ - F(\vec{k}_{\bot},\vec{q}_{1})|^2 \ \ ,\
   |\vec{q}|=|\vec{k}|=q~.
\end{split}
\end{equation}
The second term in (\ref{diff11b}) is turned to zero, owing to the
completeness of the system of basic functions $\xi
(\vec{k}_{\bot},\vec{\mu })$ \cite{part1} :
$$ \int    \mathrm{d}\Omega_{\vec{\mu}}\   \bar{\xi }(\vec{q}_{\bot},\vec{\mu })\ \xi (\vec{k}_{\bot},\vec{\mu })\sim
\delta (\vec{k}_{\bot} - \vec{q}_{\bot})~.$$ Finally we obtain:
\begin{equation}\label{diff12a}
  N=\frac{1}{(2\pi)^2} \mathrm{Re}
 \sum \limits_{\epsilon=\pm 1}\int \mathrm{d}\vec{q}_{1}\ q^2 dq \
   \mathrm{d}\Omega_{\vec{\mu}}\  \mathrm{d}\Omega_{\vec{q}} \
   \mathrm{d}\Omega_{\vec{k}}\ \bar{\xi }(\vec{q}_{\bot},\vec{\mu })\ \xi (\vec{k}_{\bot},\vec{\mu
   })|F(\vec{q}_{\bot},\vec{q}_{1})|^2~.
\end{equation}

Such representation for $N$  is universal in the meaning that it
follows both differential cross-sections in terms of
$\vec{q}_{\bot}$ and also in terms of $\vec{\mu}$. This allows
mark out singular factors in $N$, connected with infinite time and
volume. Turning to the matrix elements from $\hat{F}$ to $\hat{A}$
and integrating over momentum $\vec{q}_{1}$, we get:
\begin{equation}\label{diff15}
\begin{split}
N&= \frac{V T}{(2\pi)^6} \ \mathrm{Re} \sum \limits_{\epsilon=\pm
1}\int \ q^2 dq \
   \mathrm{d}\Omega_{\vec{\mu}}\  \mathrm{d}\Omega_{\vec{q}} \
   \mathrm{d}\Omega_{\vec{k}}\ \bar{\xi }(\vec{q}_{\bot},\vec{\mu })\ \xi (\vec{k}_{\bot},\vec{\mu })\times \\
   \times & \delta (E_{q}+\sqrt{(\vec{q}-\vec{P})^2+m_{D}^{2}} -P^0)\ |A^{(\epsilon )}(\vec{q})|^2
~.\end{split}
\end{equation}
If we integrate this correlation over the parameter $\vec{\mu}$
and momentum $q$ then we automatically obtain relation
(\ref{diff8a}). But we will integrate over momentum
$\vec{k}_{\bot}$. So we have:
\begin{equation}\label{diff16}
\begin{split}
N&= \frac{V T}{(2\pi)^6} \ \mathrm{Re}  \sum \limits_{\epsilon=\pm
1} \int \ q^2 dq \
   \mathrm{d}\Omega_{\vec{\mu}}\  \mathrm{d}\Omega_{\vec{q}} \
   \ \kappa (\mu) \ \bar{\xi } (\vec{q}_{\bot},\vec{\mu })\ \times \\
   \times & \delta (E_{q}+\sqrt{(\vec{q}-\vec{P})^2+m_{D}^{2}} -P^0)\ |A^{(\epsilon
   )}(\vec{q})|^2~,
\end{split}
\end{equation}
where \cite{Gradshtein}
\begin{equation}\label{diff17}
\kappa (\mu) =\int \xi (\vec{k}_{\bot},\vec{\mu })\
\mathrm{d}\Omega_{\vec{k}}=\frac{2\pi ^2}{ch(\pi \mu
)}\frac{\sqrt{\pi}}{|\Gamma (\frac{i\mu
}{2}+\frac{3}{4})|^2}~=~\sqrt{\pi}\Big|\Gamma(\frac{i\mu}{2}+\frac{1}{4})\Big|^2~.
\end{equation}

Further integrating over $q$ in relation (\ref{diff16}) we obtain:
\begin{equation}\label{diff18}
N= \frac{V T}{(2\pi)^6} \ \mathrm{Re}  \sum \limits_{\epsilon=\pm
1} \ \int \
   \mathrm{d}\Omega_{\vec{\mu}}\  \mathrm{d}\Omega_{\vec{q}} \
   \ \kappa (\mu)\ \bar{\xi } (\vec{q}_{\bot},\vec{\mu })\ q^2 \lambda (\vec{q}_{\bot})\ |A^{(\epsilon )}(\vec{q})|^2
   ~~, q=\tilde{q}~.
\end{equation}
Taking into account (\ref{diff8a}) we can rewrite this expression
in the following form:
\begin{equation}\label{19}
  N= \frac{1}{(2\pi)^2} \ \mathrm{Re}  \sum \limits_{\epsilon=\pm 1} \ \int  \
   \mathrm{d}\Omega_{\vec{\mu}}\  \mathrm{d}\Omega_{\vec{q}} \
   \ \kappa (\mu)\ \bar{\xi } (\vec{q}_{\bot},\vec{\mu })\ \frac{dN^{(\epsilon )}}{\mathrm{d}\Omega_{\vec{q}}}\
   ,\ q=\tilde{q}~.
\end{equation}
This relation for the total number of events $N$ is right in any
frame of particles $A$ and $B$. Crossing from $N$ to differential
distribution demands fixation of frame. If the initial state set
in c.m.f, then $\tilde q$ don't depend of the scattering angle.
And in this case we have for the differential distribution on
$\vec \mu$:
\begin{equation}\label{diff20}
  \frac{dN^{\pm}}{\mathrm{d}\Omega_{\vec{\mu}}}= \frac{1}{(2\pi)^2} \ \kappa (\mu)\ Re \ \int  \
   \  \mathrm{d}\Omega_{\vec{q}} \
   \ \bar{\xi } (\vec{q}_{\bot},\vec{\mu })\ \frac{dN^{\pm}}{\mathrm{d}\Omega_{\vec{q}}}\ ,\
   q=\tilde{q}~,
\end{equation}
where
$$  \tilde{q}^2 = \frac{(s + m_{C}^2 - m_{D}^2)^2}{4s} - m_{C}^2 ~. $$
There are $q=\tilde{q}$ everywhere below (reaction surface).

Using relations between $\mu$ and $b$
\begin{equation}\label{b}
  \mu = (b^2 q^2 - \frac{1}{4})^{1/2}\ ,\ \mathrm{d}\Omega_{\vec{\mu}}= q^2 \tanh(\pi \mu )d \vec{b}\
,\ d \vec{\mu} =\mu \ d\mu \ d \varphi ,\ d \vec{b} =b\  db\ d
\varphi~,
\end{equation}
 we get a distribution on
$\vec{b}$. So we have for differential cross-sections:
\begin{equation}\label{diff21}
   \begin{split}
  \frac{d\sigma ^{\pm}}{\mathrm{d}\Omega_{\vec{\mu}}}&= \frac{1}{(2\pi)^2} \ \kappa (\mu)\ \mathrm{Re} \ \int  \
   \  \mathrm{d}\Omega_{\vec{q}} \
   \ \bar{\xi } (\vec{q}_{\bot},\vec{\mu })\ \frac{d\sigma ^{\pm}}{\mathrm{d}\Omega_{\vec{q}}}~,\\
\frac{d\sigma ^{\pm}}{\mathrm{d}\Omega_{\vec{q}}}&=(2\pi)^2  \
\frac{q^2 \lambda (\vec{q}_{\bot} )}{|\vec{u}|}\ \
  |A^{(\epsilon )}(\vec{q})|^2 ~,\\
  \lambda (\vec{q}_{\bot})&= \frac{q E_{q} E_{D}}{q^2 P^0 -(\vec{q}\cdot
\vec{P})E_{q}}~,\\
E_{D}&=\sqrt{(\vec{q}-\vec{P})^2+m_{D}^{2}}~.
   \end{split}
\end{equation}
The matrix element $\hat{A}$ is connected with $\hat{S}$-matrix by
relation $<f|S|in>=<f|in>+i \delta
^{(4)}(q_{f}-q_{in})<f|\hat{A}|in>$~.

Now, let us integrate $\frac{d\sigma
^{\pm}}{\mathrm{d}\Omega_{\vec{\mu}}}$  (\ref{diff21}) over the
direction of vector $\vec{\mu}$:
\begin{equation}\label{diff25}
\frac{d\sigma ^{\pm}}{d\mu } =\frac{1}{2\pi }\ \mu th(\pi \mu
)\kappa (\mu)\int
\mathrm{d}\Omega_{\vec{q}}\left\{\frac{q}{\sqrt{q^2
-q_{\bot}^2}}P_{-\frac{1}{2}+i\mu }(\frac{q}{\sqrt{q^2
-q_{\bot}^2}})\frac{d\sigma ^{\pm}}{\mathrm{d}\Omega_{\vec{q}}}
\right \}~.
\end{equation}
Here we used an integral representation of cone function
\cite{Bateman}:
\begin{equation}\label{diff26}
 \int \limits _{0}^{2\pi} d \varphi \ (u_0 -\sqrt{u_0 ^2 -1}\cos(\varphi-\theta))^{-1/2 +\mathrm{i}\mu}=2\pi P_{-1/2
 + \mathrm{i}\mu}(u_0)~.
\end{equation}
Subsequent  transformation of differential cross-section on
$\vec{\mu}$ connect with turning to hyperbolic variables:
\begin{equation}\label{algebra16}
  u=(u_0 ,u_1 ,u_2)\ \ ,\  u_0 =\frac{q}{\sqrt{q^2-q_{\bot}^2}} \ ,\ \vec{u} =\frac{\vec{q}_{\bot}}{\sqrt{q^2-q_{\bot}^2}}\ \ ,\ \ u^2=u_0 ^2-u_1 ^2-u_2 ^2=1
\end{equation}

In this variables the differential volume is:
\begin{equation}\label{al16a}
   \mathrm{d}\Omega_{\vec{q}}=\frac{d\vec{q}_{\bot}}{q\sqrt{q^2-q_{\bot}^2}}=\frac{d\vec{u}}{u_0^3 }=\frac{du_0 d\varphi}{u_0 ^2}
\end{equation}
where $\varphi $ is the azimuth angle of vector $\vec{q}_{\bot}$ .

Taking into account that the differential cross-section
$\frac{d\sigma ^{\pm}}{\mathrm{d}\Omega_{\vec{q}}}$ does not
depend on $\varphi $ we obtain:
\begin{equation}\label{diff25a}
\frac{d\sigma ^{\pm}}{d\mu } = \mu \tanh(\pi \mu )\kappa (\mu)\int
\limits _{1}^{\infty}
 \frac{du_{0}}{u_{0}} P_{-\frac{1}{2}+i\mu }(u_{0})\left ( \frac{d\sigma
^{\pm}}{\mathrm{d}\Omega_{\vec{q}}} \right )~,
\end{equation}
where the angular part of the cross-section in the right part of
integral is expressed through the variable  $u_0$. Finally we
represent the differential cross-section on  $\mu$ through the
differential cross-section on scattering angle. Let $\theta$  and
$\varphi$ be axial and azimuth angles of the momentum $\vec{q}$.
Let us turn to integrating over this angles in the expression
(\ref{diff25}). For this we notice that for arbitrary function
$f(\vec{q})$ the following integral relation is right:
\begin{equation}\label{diff22}
  \int f(\vec{q})\  \mathrm{d}\vec{q}= \int q^2 dq \
  \mathrm{d}\Omega \ f(\vec{q})=\sum\limits _{\epsilon=\pm 1}\int q^2 dq \
  \mathrm{d}\Omega_{\vec{q}}\ f(\vec{q}_{\bot},\epsilon
  \sqrt{q^2-q_{\bot}^2})~,
\end{equation}
$$\mathrm{d}\Omega =\sin \theta d \theta d\varphi \ \ , \ \mathrm{d}\Omega_{\vec{q}} =
\frac{1}{q\sqrt{q^2-q^2_\bot }}\ \mathrm{d}\vec{q_\bot}~. $$ From
this follows that:
\begin{equation}\label{diff23}
   \int   \mathrm{d}\Omega \ f(\vec{q})=\sum\limits _{\epsilon=\pm 1}\int  \
  \mathrm{d}\Omega_{\vec{q}}\ f(\vec{q}_{\bot},\epsilon
  \sqrt{q^2-q_{\bot}^2})~,
\end{equation}
or in terms of $\theta $ and $\varphi $ angles
\begin{equation}\label{diff24}
\begin{split}
\int  \mathrm{d}\Omega_{\vec{q}} \ f(\vec{q}_{\bot},q_3 =+\sqrt{q^2 -q_{\bot}^2})&=\int \limits
_{0} ^{1} dz \int \limits _{0} ^{2 \pi} d\varphi f(\vec{q}_{\bot},q_3 =qz)~,\\
\int  \mathrm{d}\Omega_{\vec{q}} \ f(\vec{q}_{\bot},q_3
=-\sqrt{q^2 -q_{\bot}^2})&=\int \limits _{-1} ^{0} dz \int \limits
_{0} ^{2 \pi} d\varphi f(\vec{q}_{\bot},q_3 =qz)~,
  \end{split}
\end{equation}
$$  z=\cos \theta \ ,\ f(\vec{q}_{\bot},q_3 =qz)\equiv f(\vec{q})~.$$
So relations between variables are
$$ qz=
  \begin{cases}
    +\sqrt{q^2-q_{\bot}^2} & \text{forward half-sphere}, \\
    -\sqrt{q^2-q_{\bot}^2} & \text{backward half-sphere}.
  \end{cases}
  \ ,\ q_{\bot}=q\sqrt{1-z^2}~.
$$
We turn to integrating over angles in the integral (\ref{diff25}).
We obtain:
\begin{equation}\label{diff27}
\begin{split}
\frac{d\sigma ^{+}}{d\mu } &=\frac{1}{2\pi }\ \mu \tanh(\pi \mu
)\kappa (\mu) \int \limits _{0}
^{1} \frac{dz}{z}P_{- 1/2+i\mu}(\frac{1}{z})\frac{d\sigma }{dz}~,\\
\frac{d\sigma ^{-}}{d\mu } &=\frac{1}{2\pi }\ \mu \tanh(\pi \mu
)\kappa (\mu) \int \limits _{-1} ^{0} \frac{dz}{|z|}P_{-
1/2+i\mu}(\frac{1}{|z|})\frac{d\sigma }{dz}~.
\end{split}
\end{equation}
where $\frac{d\sigma }{dz}$ is the differential cross-section on
the cosine of scattering angle of particle $C$ in the
$A+B\rightarrow C+D$ process. Here is taken into account that
$\frac{d\sigma ^{\pm}}{\mathrm{d}\Omega_{\vec{q}}}$ do not depend
of variable $\varphi$, therefore
$$\frac{d\sigma ^{\pm}}{\mathrm{d}\Omega_{\vec{q}}}= \frac{1}{2 \pi} \frac{d\sigma}{dz} ~. $$
Unifying relations (\ref{diff17}) , ( \ref{diff21}) , (
\ref{diff27}) and taking into account that:
$$ \frac{1}{(2\pi)^2 }\int \kappa (\mu) \bar{\xi } (\vec{q}_{\bot},\vec{\mu
})\mathrm{d}\Omega_{\vec{\mu}}=1~, $$ we obtain the norm of the
differential cross-section on the total cross-section
$\sigma(AB\rightarrow CD) $
\begin{equation}\label{norma}
 \sum \limits_{\epsilon=\pm 1} \int \limits _{0} ^{\infty}\left( \frac{d\sigma ^{(\epsilon )}}{d\mu}\right) d\mu =
 \int \limits _{-1} ^{1}\left( \frac{d\sigma}{dz} \right ) dz =\frac{N}{TVn_1 n_2 |\vec{u}|}=\sigma(AB\rightarrow
 CD)~.
\end{equation}

As follows from (\ref{diff21}), in contrast to $\frac{d\sigma
^{\pm}}{\mathrm{d}\Omega_{\vec{q}}}$  the differential
cross-section $\frac{d\sigma
^{\pm}}{\mathrm{d}\Omega_{\vec{\mu}}}$ is  not positively sign
determined on all $\mu$ interval. Contribution of negative value
area of $\frac{d\sigma ^{\pm}}{\mathrm{d}\Omega_{\vec{\mu}}}$
reduces effectively to decreasing of the total event number of
particle $C$ creation. So, this spatial area we can interpret as
area where taking place an absorption of particles $C$. But the
total number of asymptotic states with define $\mu$ regulates by
the relation (\ref{norma}).

\section{Application to simple models $\frac{d\sigma}{d\Omega}$ in c.m.f.}

As an example we discuss an one-particle exchange in $t$-channel,
for elastic $A+B\rightarrow A+B$ scattering. Corresponding cross
section as function of $u_0$ has a polar view:
\begin{equation}\label{Elastic}
\frac{d\sigma}{d\Omega}
~=~N_0\frac{\alpha^2}{(t-M^2)^2}~=~\frac{\alpha^2 N_0}{(2 q^2)^2}
\frac{1}{(z_0-z)^2}~=~\frac{\alpha^2 N_0}{(2 z_0
q^2)^2}\frac{u_0^2}{(u_0-\frac{\varepsilon}{ z_0})^2}~,
\end{equation}
here
$$z_0~=~1+\frac{2M^2s}{\lambda(s,m_A^2,m_B^2)}~=~1+\frac{M^2}{2q^2}~,$$
$M$ -- mass of exchange particle,
$\lambda(x,y,z)~=~(x^2+y^2+z^2-2xy-2xy-2yz$) -- wellknown the
triangle function, $\alpha$ -- the coupling constant,
$$N_0~=~(2\pi)^2\frac{q^2 \tilde\lambda(\vec q_\perp)}{|\vec u|}
~=~(2\pi)^2\frac{(s^2-(m_1^2-m_2^2)^2)^2}{16 s^3}~.$$

In this case for the cross-section $\sigma^\pm$ we obtain:
\begin{equation}
\sigma^{\pm} ~=~\frac{\alpha^2 N_0}{(2q^2)^2}
\frac{2\pi}{z_0(z_0\mp 1)}~,
\end{equation}
where
\begin{equation}\label{sigmaPM}
\sigma^{(\varepsilon)}~=~ \int \limits _{0} ^{\infty}\left(
\frac{d\sigma ^{(\epsilon )}}{d\mu}\right) d\mu =\left\{
\begin{array}{c}
 \int \limits _{0} ^{1}\left( \frac{d\sigma}{dz} \right )
 dz~,~~\varepsilon~=~1\\~\\
\int \limits _{-1} ^{0}\left( \frac{d\sigma}{dz} \right )
dz~,~~\varepsilon~=~-1
\end{array}\right.~~.
\end{equation}

 The integral
(\ref{diff25a}) with such cross section is computed analytically.
The normalize distribution on  nearest approach parameter takes a
view:
\begin{equation}\label{ElasticCS}
\frac{1}{\sigma^\varepsilon}\frac{d\sigma ^{\varepsilon}}{d
b}~=~\frac{q^2 b}{2} \frac{\tanh(\pi \mu )\kappa
(\mu)}{\cosh(\mu\pi)}\frac{z_0-\varepsilon}{z_0}\Bigg[P_{i\mu-1/2}\Big(\frac{-\varepsilon}{z_0}\Big)+\frac{\varepsilon
}{\sqrt{z_0^2-1}}P_{i\mu-1/2}^1\Big(\frac{-\varepsilon}{z_0}\Big)\Bigg]~,
\end{equation}
where $\varepsilon~=~\pm1$, and $P_{i\mu-1/2}^1(x)$ -- associated
cones function. The plot of (\ref{ElasticCS}) is shown on Fig.1a.
 \begin{figure}[t] \label{f:115}
\begin{center}
a.\includegraphics*[scale=0.45] {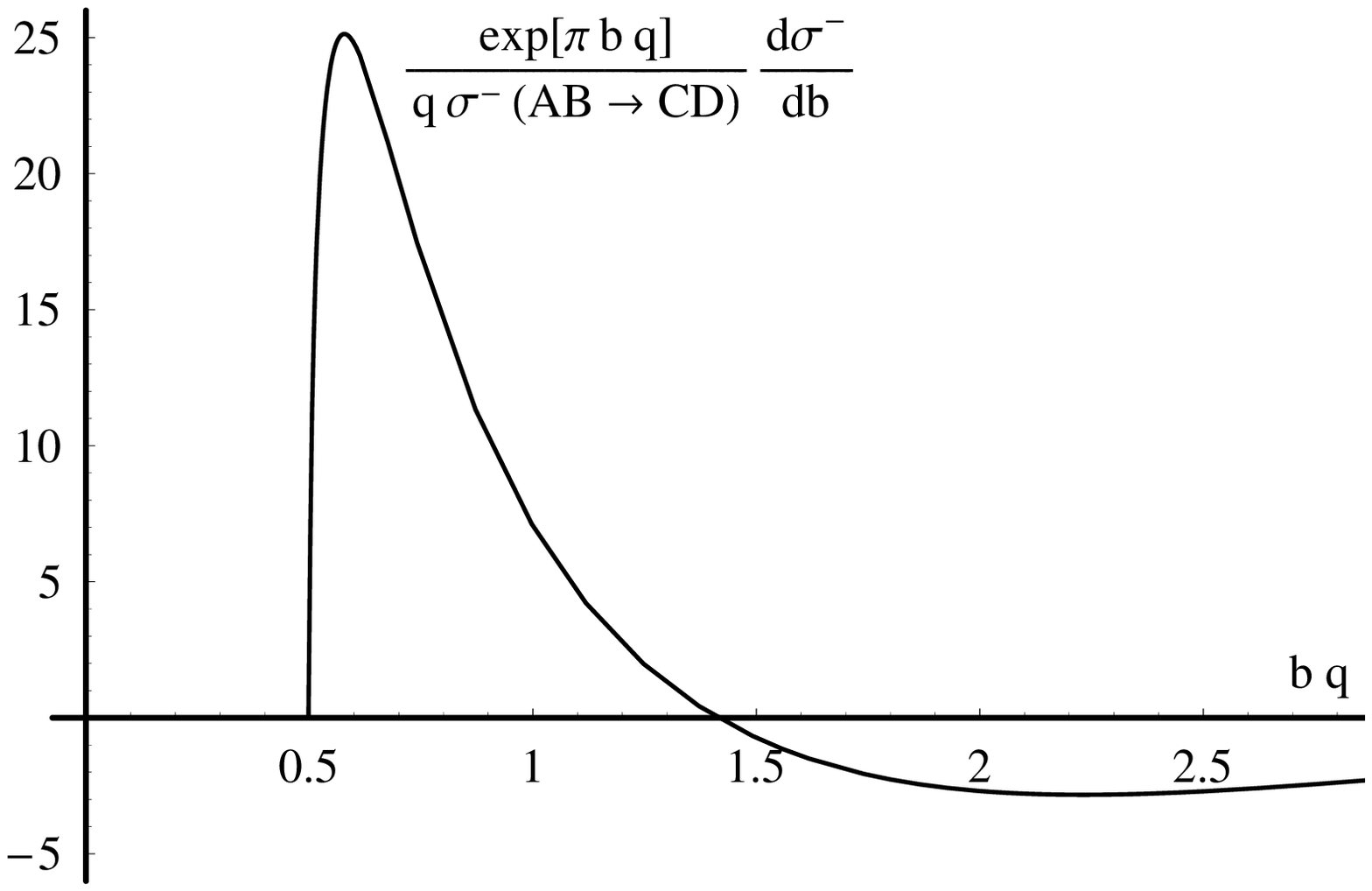}
b.\includegraphics*[scale=0.4] {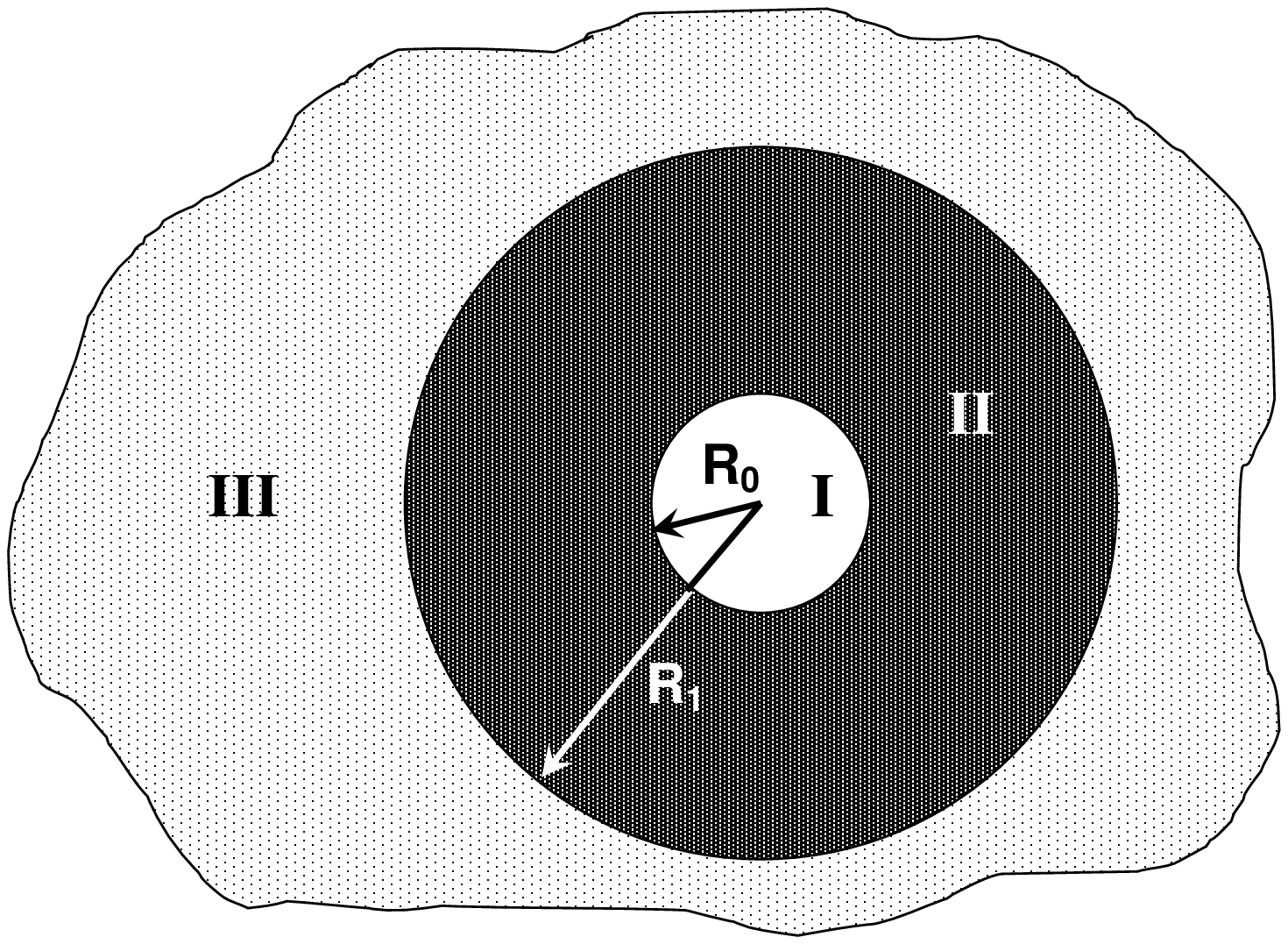}
\caption{a)Distribution function on $b$ in the model of one
particle exchange. $bq<1/2$ -- area forbidden by uncertainty
relation, $1/2<bq\lesssim\sqrt 2$ -- area of creation particles
$C$, $bq\gtrsim\sqrt{2}$ -- area of absorption particles $C$.
b)The zone structure of plain $\vec b$ in the model of one
particle exchange. Zone I -- area forbidden by uncertainty
relation, zone II --area of creation particles $C$, zone III --
area of absorption particles $C$.} $z_0 =1.001$
\end{center}
\end{figure}

Since $z_0\geqslant 1$ so the argument of cones functions belong
to segment $[-1,1]$. The cones function is positive with such
argument (and associated cones functions with $m>0$ too). So the
cross-section is positive defined on $\varepsilon=1$. On
$\varepsilon=-1$ the $\vec{b}$-plane divides into zones with
radiuses $R_0$ , $R_1$ and $R_2$ Fig.1b.

Here $R^2_{0}=\hbar ^2 /4 q^2$ defines a border of forbidden area,
where Geizenberg uncertainty relation is broken (phase space of
particle $C$ is less then allowable). Area $R_0 < b < R_1$ defines
spatial area of particle $C$ creation. And area $ b
> R_1 $ is area where absorption of particles is taken place relating to equality (\ref{norma}).
At that, if we symbolically take density of distribution $I(b)
\sim 1$ in area $R_0 < b < R_1 $, when density in area $b
> R_1 $ would be $\sim 10^{-5}$. It is made conditional upon that exponential
decrease of $I(b)$ in the backward sphere determines by radius $R
\sim 1/ q $  i.e. $I(b)\sim \exp(-\pi bq)$ at $bq \gg 1$ .

The left border of $R_2$-zone is defined by zero of expression
(\ref{ElasticCS}). In the area of small values of parameter
$\frac{M^2}{s}$ the cross-section (\ref{ElasticCS}) turn into zero
on
$$\mu~=~\mu_0~=~\frac{\sqrt{7}}{2}+2\frac{2}{\sqrt{7}}\epsilon+\frac{2}{21\sqrt{7}}\epsilon^2+O(\epsilon^3)
~,~~~~~~~~\epsilon~=~\frac{2M^2s}{\lambda(s,m_1^2,m_2^2)}\sim\frac{M^2}{s}~.$$

In the area $\frac{M^2}{s}\ll 1$ the cross-section takes a view:
\begin{equation}
\frac{1}{\sigma^-}\frac{d\sigma ^{-}}{d b}~=~q^2 b \frac{\tanh(\pi
\mu )\kappa
(\mu)}{\cosh(\mu\pi)}\Bigg[\frac{7-4\mu^2}{8}+O(\frac{M^2}{s})\Bigg]~,
\end{equation}
\begin{equation}
\frac{1}{\sigma^+}\frac{d\sigma ^{+}}{d b}~=~q^2 b \frac{\tanh(\pi
\mu )\kappa
(\mu)}{\cosh(\mu\pi)}\Bigg[\frac{\cosh(\mu\pi)}{\pi}+O(\frac{M^2}{s})\Bigg]~.
\end{equation}

\section{Connection between $<b^2>$ and the cross-section on transverse momentum of particle $C$ in C.M.F of $A$ and $B$ particles}

Let us calculate the an average value of  nearest approach
parameter square $<b^2_\pm>$. We have by definition:
\begin{equation}\label{<mu^2>}
<\mu^2_\pm>~=~\frac{1}{\sigma^\pm}\int_0^\infty \mu^2
\frac{d\sigma^\pm}{d \mu}d\mu~,
\end{equation}
where $\sigma^\pm$ was defined in (\ref{sigmaPM}). So:
\begin{equation}\label{<mu^2>2}
<b^2_\pm>~=~\frac{1}{q^2}\Big(<\mu^2_\pm>+\frac{1}{4}\Big)~,~~~q~=~\tilde
q~.
\end{equation}
Substituting the representation for $\frac{d\sigma^\pm}{d\mu}$
from (\ref{diff25a}), into (\ref{<mu^2>}) and taking into account
that (compare with (\ref{diff17}) and (\ref{diff26})):
$$\kappa(\mu)~=~2\pi \int_1^\infty P_{i\mu-1/2}(x)\frac{dx}{x}~,$$
we obtain
\begin{equation}\label{<b^2>}
<b^2_\pm>~=~\frac{2\pi}{\tilde q^2 \sigma^\pm}\int
(\mu^2+\frac{1}{4})P_{i\mu-1/2}(x)P_{i\mu-1/2}(u_0)~\frac{d\sigma^\pm}{d\Omega}\frac{dx}{x}\frac{du_0}{u_0}
d\Omega_\mu~.
\end{equation}

The differential equation on cones functions $P_{i\mu-1/2}(u_0)$
with argument $u_0$ can be represented in following form
\cite{Bateman}:
$$(\mu^2+\frac{1}{4})P_{i\mu-1/2}(u_0)~=~
\Big(-(u_0^2-1)\frac{d^2}{du_0^2}-2u_0\frac{d}{du_0}\Big)P_{i\mu-1/2}(u_0)~.$$
The left part of this expression went as underintegral factor into
relation (\ref{<b^2>}) and its substitution allows us integrate
over $\mu$. From the completeness relation of cones functions
follow  \cite{part1}:
\begin{equation}
\int_{0}^\infty
P_{i\mu-1/2}\big(x\big)P_{i\mu-1/2}\big(u_0\big)~d\Omega_\mu~=~
\delta(x-u_0)~.
\end{equation}
Next, we can integrate over variable $x$. After some
transformations we obtain:
\begin{equation}
<b^2_\pm>~=~\frac{2\pi}{\tilde q^2 \sigma^\pm}\int
\frac{2}{u_0^3}~\frac{d\sigma^\pm}{d\Omega}\frac{du_0}{u_0}~.
\end{equation}
Crossing in this relation from the variable $u_0$ to the
scattering angle $\theta$, we obtain:
\begin{equation}
<b^2_\varepsilon>~=~\left\{ \begin{array}{c} \displaystyle{
\frac{2}{q^2 \sigma^+}\int_0^1
\cos^2\theta\frac{d\sigma^+}{d\cos\theta}d\cos\theta
~,~~\varepsilon~=~1}
\\
\displaystyle{ \frac{2}{q^2 \sigma^-}\int_{-1}^0
\cos^2\theta\frac{d\sigma^-}{d\cos\theta}d\cos\theta~,~~\varepsilon~=~-1}
\end{array}
 \right.~~.
\end{equation}
So, finally we have:
\begin{equation}\label{<b^2>2}
<b^2_\pm>~=~ \frac{8~
s}{\lambda(s,m^2_C,m^2_D)}<\cos^2\theta_\pm>~.
\end{equation}
where
$$<\cos^2\theta_\varepsilon>~=~\left\{ \begin{array}{c} \displaystyle{
\frac{1}{\sigma^+}\int_0^1 z^2\frac{d\sigma}{dz}dz
~,~~\varepsilon~=~1}
\\
\displaystyle{ \frac{1}{\sigma^-}\int_{-1}^0
z^2\frac{d\sigma}{dz}dz~,~~\varepsilon~=~-1}
\end{array}
 \right.~~,~~~z~=~\cos\theta~, $$
and $<b^2_\pm>$ is defined by relation (\ref{<mu^2>}).

It follows:
\begin{equation}\label{<mu^2>3}
<\mu^2_\pm>~=~2<\cos^2\theta_\pm>-\frac{1}{4}~.
\end{equation}

The parameter $\mu$ is real number, so $\mu^2>0$. From it and
(\ref{<mu^2>3}) follows an important physical inequality:
\begin{equation}
<\cos^2\theta_\pm>~ \geqslant~\frac{1}{8}~.
\end{equation}

Let us discuss the nature of this inequality. It follows from the
reality of parameter $\mu~=~\sqrt{b^2q^2/\hbar^2-1/4}$, where
$\hbar$ -- Plank constant. Here $b^2$ is an eigen value of
Kazimir's operator on $SO(2,1)$-group \cite{part1}. Spectrum of
this operator in Hilbert space of states satisfies condition
$b^2q^2\geqslant \hbar^2/4$, and that is provide a reality of
parameter $\mu$. By itself this inequality has an quantum nature
and show us the fact that particle $C$ can not creates in phase
space less when it allows by Geizenberg's uncertainty relation.

We notice on the fact that relation (\ref{<b^2>2}) between
$<b^2_\pm>$ and $<\cos\theta_\pm^2>$ right for any process
$A+B\rightarrow C+D$ in center of mass frame $A$ and $B$
particles.

In the model with one-particle exchange (\ref{Elastic}) it is easy
to obtain that
$$<\cos^2\theta_\pm>~=~\pm 2 z_0^2(z_0\mp 1)\ln\Big(\frac{z_0\mp
1}{z_0}\Big)\mp z_0 + 2 z^2_0~.
$$
So it follows:
$$<b^2_\pm>~=~\frac{1}{q^2} \Bigg[\pm 4 z_0^2(z_0\mp 1)\ln\Big(\frac{z_0\mp
1}{z_0}\Big)\mp 2z_0 + 4 z^2_0\Bigg]~.
$$

Let us analyze this expressions as functions of parameter
$z_0=1+M^2/2q^2$ (relation (\ref{Elastic})). The analyze shows
that $<\cos^2\theta_\pm>$ changes in borders:
$$\frac{1}{3}~<~<\cos^2\theta_+>~\leqslant~1~~,$$
$$0.23~\lesssim~<\cos^2\theta_->~<~\frac{1}{3}~~,$$
when $z_0$ changes in interval from $1$ to $\infty$.

\section{Differential cross section on  nearest approach parameter $\vec b$ in l.f.}

As it was note in the previous section, in the labor frame of $B$
particle energy and momentum of $C$ particle depend of scattering
angle (\ref{L.S.C.}). Consequence of this fact is appearance on
the reaction plane $q=\tilde q$ as the parameter $\mu$ dependence
from the scattering angle, if we use relation (\ref{b}) for the
transition to the parameter $b$. $SO_\mu(2,1)$ algebra implements
on basis functions, for which the parameter $\mu$ is natural
variable. Completeness and orthogonality of basis are appeared in
terms of it. If we take $\mu$ in relation (\ref{19}) as an
independent parameter, then differential cross section
$\frac{d\sigma }{d \mu}$ is defined unambiguously and doesn't
depend of frame. But at the same time the physical meaning of this
parameter isn't clear, only as phase space measure $(b,q_{\bot})$.
In c.m.f. transition in the integral (\ref{19}) from $\mu$ to $b$
does not change anything considerably, and differential
distributions on $\mu$ and on $b$ are equivalent. But in l.f. the
differential distribution changes radically. Let us cross in the
integral (\ref{19}) to the variable $\vec{b}$ using relations
(\ref{b}). Than we have:
\begin{equation}\label{b1}
  N= \frac{1}{(2\pi)^2} \ \mathrm{Re}  \sum \limits_{\epsilon=\pm 1} \ \int  d \vec{b} \int \
    \mathrm{d}\Omega_{\vec{q}} \ \vartheta [b - R_{0}(\tilde{q})] q^2 \tanh(\pi \mu )
   \ \kappa (\mu)\ \bar{\xi } (\vec{q}_{\bot},\vec{\mu })\ \frac{dN^{(\epsilon
   )}}{\mathrm{d}\Omega_{\vec{q}}}~,
\end{equation}
here $$ \mu = (b^2 q^2 - \frac{1}{4})^{1/2}\ ,\ R_{0}(\tilde{q}) =
\frac{\hbar}{2\tilde{q}}\  , \ \vartheta (x)-\textrm{discontinuous
function}\ ,\ q=\tilde{q }\ , $$
 and \ \ $\tilde{q }= p_{C}$ defines in (\ref{L.S.C.})

From this it follows for the differential cross section on $\vec
b$:
\begin{equation}\label{b2}
  \frac{d\sigma ^{(\epsilon )}}{d \vec{b}}= \frac{1}{(2\pi)^2} \ \mathrm{Re}  \sum \limits_{\epsilon=\pm 1}  \int \
    \mathrm{d}\Omega_{\vec{q}} \ \vartheta [b - R_{0}(\tilde{q})] q^2 th(\pi \mu )
   \ \kappa (\mu)\ \bar{\xi } (\vec{q}_{\bot},\vec{\mu })\ \frac{d\sigma^{(\epsilon
   )}}{\mathrm{d}\Omega_{\vec{q}}}~.
\end{equation}
Integrating left and right part over the direction of $\vec b$
vector and crossing to angle variables $\theta$ and $\varphi$ of
$\vec{q}_{\bot}$ vector we obtain the result, which is analogous
to relations (\ref{diff27}):
\begin{equation}\label{b3}
\begin{split}
\frac{d\sigma ^{+}}{d b} &=\frac{b}{2\pi } \int \limits _{0} ^{1}
\frac{dz}{z}\vartheta [b - R_{0}(\tilde{q})] q^2 \tanh(\pi \mu )
   \ \kappa (\mu)   \ P_{- 1/2+i\mu}(\frac{1}{z})\ \frac{d\sigma }{dz}~, \\
\frac{d\sigma ^{-}}{d b} &= \frac{b}{2\pi } \int \limits _{-1}
^{0} \frac{dz}{|z|}\vartheta [b - R_{0}(\tilde{q})] q^2 \tanh(\pi
\mu )
   \ \kappa (\mu)   \ P_{- 1/2+i\mu}(\frac{1}{|z|})\ \frac{d\sigma
   }{dz}~.
\end{split}
\end{equation}
From comparison of (\ref{b3}) and (\ref{diff27}) it follows that
differences between distributions on the parameter $b$ in c.m.f.
and in l.f. generates by difference of the dependence of particle
$C$ momentum from kinematic variables $s$ and $t$ for this two
cases. From the explicit expression $\tilde{q}=p_{C}$
(\ref{diff27}) it follows that it is a fluent function of angle
and with certain accuracy it is possible cross from $\tilde q$ to
some midvalue in integrals (\ref{b3}). For example, we receive
that on the backward semisphere:
$$ \tilde{q}(z) \simeq \tilde{q}(\bar{z}) \ ,\ z=\cos (\theta ) \ ,\ \bar{z}=-0.5~,$$
and it isn't depend on angle. Then the distribution (\ref{b3})
coincides with the distribution (\ref{diff27}) at
$\tilde{q}(z)\simeq \tilde{q}(\bar{z})$. More detailed structure
of particle $C$ creation area reflects in a generalized
distribution function:
\begin{equation}\label{b4}
  \frac{d^{2} \sigma }{d b\  dz} = \frac{b}{2\pi }\cdot \vartheta [b -
R_{0}(\tilde{q})] q^2 \tanh(\pi \mu )
   \ \kappa (\mu)   \ \frac{1}{|z|} P_{- 1/2+i\mu}(\frac{1}{|z|})\ \frac{d\sigma }{dz}\ ,\ -1< z
   <1~,
\end{equation}
$$q=\tilde{q}=p_{C}(z)\ ,\ \mu = (b^2 q^2 - \frac{1}{4})^{1/2}~. $$
It gives us a tomographic picture of partial integrals over $z$ to
the spatial distribution $\frac{d\sigma}{db}$ in the whole
interval of $b$. It is similar to Wigner function \cite{Wigner}
From it we can get expressions for the differential cross section
on $q_{\bot}$ (\ref{diff8a}), for differential cross section on
$b$ (\ref{diff27}, \ref{b3}) and norm relations (\ref{norma}).

As an illustration of represented method, let us consider process
of neutral $\pi$-meson  photoproduction  on proton. There is
experimental data on angle distribution of $\pi^0$ in the interval
$- 0.75 < z=\cos\theta< 0 $ (backward semisphere) at photon energy
$E_{\gamma }= 5\textrm{GeV}$ (Fig.3a)\cite{Anderson}.
\begin{figure}[ht]
\begin{center} \label{f:20}
a.\includegraphics*[scale=0.4] {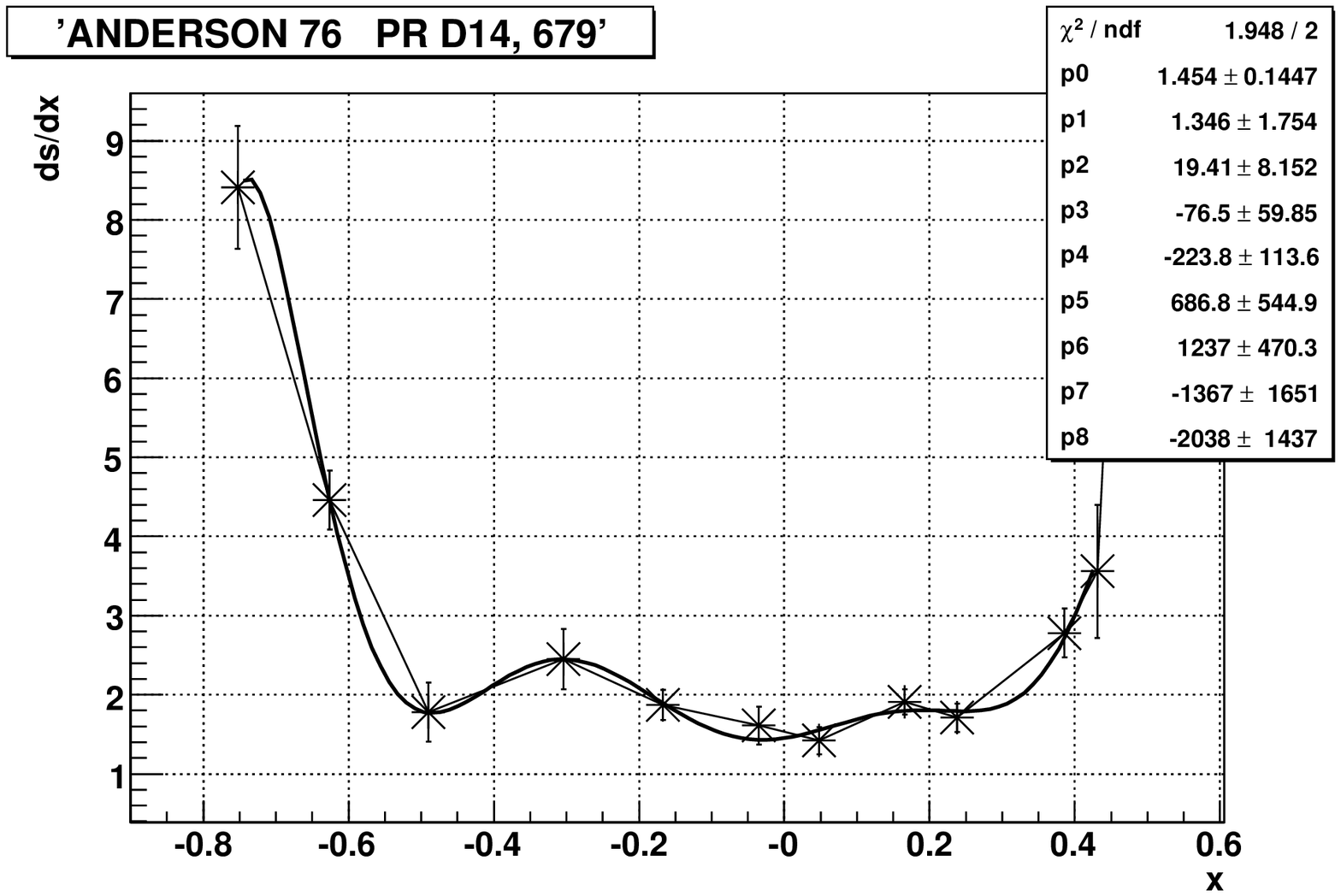}
b.\includegraphics*[scale=0.5] {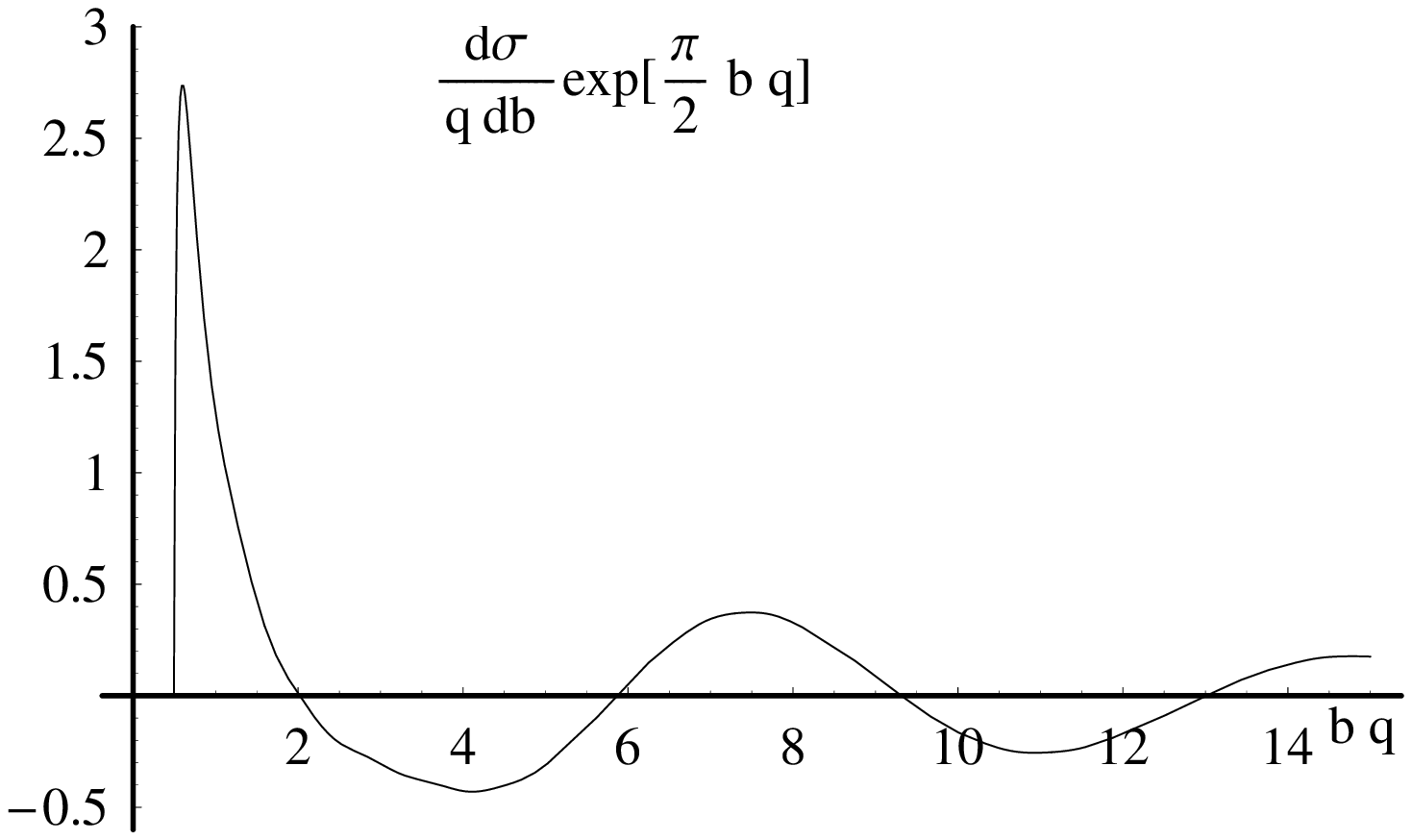} \caption{a)Differential
cross section of process $\gamma +p \rightarrow \pi ^0 +p $ , \
$x=\cos \theta $} b) Differential cross section on $b$ for process
$\gamma +p \rightarrow \pi ^0 +p $
\end{center}
\end{figure}
 At such energy and in
this angle interval the meson energy depends of angles weakly. And
we receive in relations (\ref{b4}) $\pi ^0$ momentum as its
midvalue $q \simeq 665 \textrm{MeV}$. There is the generalize
distribution function at current parameters (Fig.4).
\begin{figure}[ht]
\begin{center}
\label{f:23}
\includegraphics*[scale=0.6] {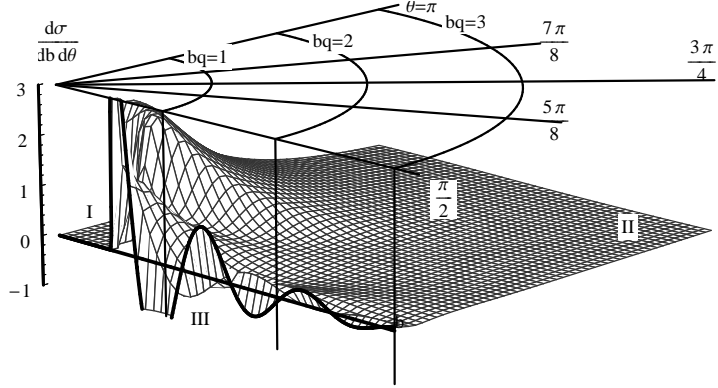}
\caption{Generalize distribution function in plane ($b,z$) for the
process $\gamma +p \rightarrow \pi ^0 +p \ ,\ E_{\gamma }=5 \
\textrm{GeV}\ ,\ q\simeq 665 \ \textrm{MeV }$ }
\end{center}
\end{figure}
The area I is defined by condition $b< R_{0}=\hbar /2 q$. This is
lower border of parameter $b$ physical values, it is consequent of
Geizenberg uncertainty relation in phase space $(b,q)$.
$R_{0}=1.498\cdot 10^{-14} \textrm{cm}$ at current values. The
area II is particle $C$ creation area. In this area the
distribution function is positive. Finally, the area III is area
where the distribution function is negative. Contribution of this
area to the total number of particle $C$ creation events is
negative. So, as it was discussed, this area is interpreted as the
particle $C$ absorption area. The total number of events in unit
time is regulate according to (\ref{norma}).


Fig.3b is distribution on $b$ integrated over available angle
interval $- 0.75 < z=\cos\theta< 0$:
$$   \frac{d\sigma}{db} = \int \limits _{-0.75}^{0} \left [ \frac{d^{2} \sigma }{d b\  dz} \right ]dz~.  $$
Evidently to the graphic, particle $C$ creation areas are also
divide to zones. Numerical calculation at current parameters gives
us the following values of zone radiuses. The first zone - (
$R_{0}<b<R_{1}$), where  $R_{1}= 5.982\cdot 10^{-14} \textrm{cm}$.
The second zone - ($R_{2}<b<R_{3}$), where  $R_{2}= 1.757\cdot
10^{-13} \textrm{cm} , R_{3}= 2.79\cdot 10^{-13} \textrm{cm}$. The
third zone - ($b>R_{4})$, where $R_{4}= 3.899  \cdot 10^{-13}
\textrm{cm}  $ and so on.

The cross section fast falls with $b$ growing. If we take the
maximum value of cross section in the first zone as unit, when the
maximum value in the first absorption zone ($bq \sim 4 $) would be
equal to $1.5 \cdot  10^{-3}$, and the maximum in the second
creation zone would be $2.8 \cdot 10^{-6}$. Here we represented
estimates of the spatial distribution starting from some midvalue
of $\pi^0$ energy.  In fact we have to do exact calculation in
(\ref{b4}) of $\mu$ angle dependence, using corresponding relation
for $\pi^0$ momentum.

\section{Conclusion}
So, in the paper it was relieved the exact relation between the
differential cross section on the detect particle momentum and the
distribution function on out going  nearest approach parameter of
this particle. This function describes a distribution of matter in
reaction area and allows understand the nature of its composite
parts on hadronic scale. This connection has a different view in
the center of mass frame of $A$ and $B$ particles and in
laboratory frame of particle $B$. On the simple model it was shown
that reaction area divides on creation and absorption areas of
particle $C$. It was received a relation between $<b^2_\pm>$ and
$<\cos^2\theta_\pm>$  in center of particles $A$ and $B$ mass
frame, where average is going on corresponding differential
cross-sections. This relation is exact and right for any
$A+B\rightarrow C+D$ processes. There is a bottom bound on
$<\cos^2\theta_\pm>$ also follows from it.

Only completeness of states $ |\vec{\mu } ,q,\epsilon > $ in one
particle Fock space was used for finding connection between cross
sections.

 It is received the generalized distribution on plane
($b,z$), and the differential cross sections and the norm follows
from that. As an illustration it was considerated a real process
of photoproduction $\pi^0$ on the proton, and it was done
valuation of the main characteristics of spatial distribution.

\section{Acknowledgements}
It is our pleasure to thank prof. I.B.Khriplovich for constructive
discussion and critical remarks. This investigation has been
supported in part by grant President of Russian Federation for
support of leading scientific schools
 (NSh -5362.2006.2) (N.B \& A.N.V), by the Russian Foundation for Fundamental
 Research (A.A.V), and
by the Heisenberg-Landau Program of JINR
 grant No. 07-02-91557 (A.A.V)

\end{document}